\documentclass[sigconf,nonacm]{acmart}

\copyrightyear{2026}
\acmYear{2026}
\setcopyright{cc}
\setcctype{by}
\acmConference[ICAIF '26]{7th ACM International Conference on AI in Finance}{November 14--17, 2026}{Milan, Italy}
\acmBooktitle{7th ACM International Conference on AI in Finance (ICAIF '26), November 14--17, 2026, Milan, Italy}
\acmDOI{10.1145/3858063.3858768}
\acmISBN{979-8-4007-3087-0/2026/11}

\makeatletter
\@ACM@balancefalse
\makeatother

\usepackage{booktabs}
\usepackage{amsmath}
\usepackage{tabularx}
\usepackage{tikz}
\usetikzlibrary{arrows.meta,positioning}
\definecolor{qbxgreen}{HTML}{1F6F54}
\definecolor{qbxlight}{HTML}{E8F2EE}
\definecolor{rejectred}{HTML}{A33A34}
\definecolor{rejectlight}{HTML}{F8ECEA}
\newtheorem{defn}{Definition}
\newtheorem{prop}{Proposition}

\newcommand{\WF}{\mathrm{WF}}
\newcommand{\phieff}{\varphi}
\newcommand{\ex}[1]{\textsf{E#1}}

\begin{document}

\title{QBX: A Compiler for Well-Formed, Auditable Alpha Programs}

\author{Yichong Bai}
\authornote{The QBX system underlying this study was completed by the authors before March 2026, prior to the first author's joining The Hong Kong Polytechnic University (PolyU). This paper provides its first peer-reviewed account, together with a formal analysis and empirical evaluation of its properties.}
\orcid{0009-0007-0835-0135}
\affiliation{%
  \institution{The Hong Kong Polytechnic University}
  \city{Hong Kong}
  \country{China}}
\email{byc@pku.org.cn}

\author{Kaigui Bian}
\orcid{0000-0003-0136-6082}
\affiliation{%
  \institution{Peking University}
  \city{Beijing}
  \country{China}}
\email{bkg@pku.edu.cn}

\renewcommand{\shortauthors}{Bai and Bian}

\authorsaddresses{}
\renewcommand{\shorttitle}{QBX: Well-Formed, Auditable Alpha Programs}
\titlenote{Author-prepared preprint. ICAIF '26, November 14--17, 2026, Milan, Italy. DOI: \url{https://doi.org/10.1145/3858063.3858768}. Licensed under Creative Commons Attribution 4.0 (CC BY 4.0).}

\begin{abstract}
Formulaic alpha miners increasingly differ in how they generate
candidates, yet each system typically couples its own representation,
evaluator, and selector into a bespoke pipeline whose results are
difficult to reproduce or compare. We present \emph{QBX}, a compiler
that defines a well-formed, auditable space of alpha programs. A
finite, versioned block registry and whole-strategy port-DAG language
define the candidate space; the compiler either rejects a proposal
with structured diagnostics or emits an evidence-carrying IR with
types, time-availability effects, executable linkage, and Merkle-style
content identities. Compiler-closed actions keep search inside the
admitted space, node identities support incremental evaluation, and
context-bound root identities govern program deduplication and trial
provenance. The availability-effect system rejects contract-visible
look-ahead while admitting legal point-in-time forecasts.

On Chinese A-share panels, evaluation-valid proposals increase from
26.8\% for free strings and 36.4\% for a grammar-only control to
97.7\% for typed programs. Incremental evaluation replays search
lineages $7.5\times$ faster with numerical parity, while the
availability checker admits zero registered-contract violations across
23{,}342 candidates. Under matched LLM-call budgets, compiler-closed
search reaches the pool-specific diagnostic frontier with substantially
fewer calls. These results concern validity, reuse, and search efficiency
within the registered vocabulary, not an increase in the information
available from the market panel.
\end{abstract}

\ccsdesc[500]{Applied computing~Economics}
\ccsdesc[300]{Software and its engineering~General programming languages}

\keywords{alpha mining, domain-specific languages, compilers, data leakage, backtest overfitting, LLM agents, reproducibility}

\maketitle

\section{Introduction}
A growing fraction of AI-in-finance research mines \emph{formulaic alphas}: symbolic expressions over market panels whose cross-sectional ranks predict returns \cite{kakushadze2016alphas}. The generator has evolved quickly---genetic programming and learned graph search \cite{cui2021alphaevolve}, reinforcement learning \cite{yu2023alphagen,ren2024riskminer}, Monte-Carlo tree search \cite{shi2025alphajungle}, and now large language models and agent stacks \cite{wang2023alphagpt,tang2025alphaagent,guo2026alphaprobe,shi2026hubble}. The measurement layer has not. Each paper ships a vertically integrated pipeline---its own candidate representation, its own evaluator, its own selection rule---and reports a headline information coefficient that is not comparable across papers, frequently not reproducible within them, and often inflated by mechanisms the pipeline cannot even represent, let alone control.

Three failure surfaces recur. \emph{Admission}: free-text candidates are malformed, ill-typed, or read the future; validity is discovered only by running a backtest, so the expensive evaluator doubles as a syntax checker and the leak detector. \emph{Identity}: the same signal reappears under renamings, algebraic aliases, and re-derivations; pipelines re-evaluate duplicates and then count them as independent discoveries. \emph{Selection}: the reported champion is the maximum over an unrecorded number of trials; without a trial ledger, multiple-testing corrections such as the deflated Sharpe ratio \cite{bailey2014deflated,harvey2015backtesting} have no defensible denominator.

This paper proposes and evaluates a different unit of progress: a \emph{substrate}, built in three stages. Stage~I is a \emph{registered strategy space}: a finite typed vocabulary of blocks and a port-graph strategy language whose membership is decided by a compiler, giving the field a definition space any miner---LLM, genetic, RL, or human---can target. Stage~II is a \emph{compiler for alpha mining}: a deterministic pipeline that rejects a candidate with a structured diagnostic or admits it with evidence-carrying intermediate representation (IR)---a topological schedule, dock-level value types, a signed \emph{time-availability} effect $\phieff$, and Merkle-style content addresses. Stage~III is the \emph{computation layer}, oriented at backtesting: node-level addresses price and reuse execution, the root address deduplicates and counts candidates, and the resulting trial ledger is the denominator object that selection statistics need---an interface we define here and whose full statistical exploitation we leave to future work (\S\ref{sec:concl}).

Because representation cannot add information to a fixed market panel, we evaluate QBX as a systems intervention under a fixed proposal budget. Admission and deduplication are CPU-priced in microseconds, whereas an LLM proposal call takes seconds, so each call can seed a compiler-closed neighborhood of legal candidates. In the reported pools, the typed arm reaches the pool-specific diagnostic frontier with ${\ge}10\times$ fewer calls on the pre-extension panel; a hybrid that combines LLM-sourced breadth with typed search depth improves the book at $K{=}20$ and $40$; and incremental identity makes lineage replay $7.5\times$ faster (\S\ref{sec:payoff}). Admission is therefore orders of magnitude cheaper than LLM proposal or validation, but the comparison is conditional on the tested registry, candidate pools, and market panels.
\textbf{Contributions.}
(1)~\emph{A compiler-defined space of well-formed alpha programs}
(\S\ref{sec:space}--\S\ref{sec:compiler}).
We introduce a finite, versioned block registry and a whole-strategy
port-DAG language whose admitted space $S_{R,c}$ consists exactly of
programs accepted by the QBX compiler. Admission checks completeness,
registered binding, acyclicity, dock types, information availability,
semantic intent, and executable linkage. Compiler-closed graph actions
allow LLM, genetic, RL, or human proposers to navigate the admitted
space while returning only well-formed successors.

(2)~\emph{Evidence-carrying compilation for execution and audit}
(\S\ref{sec:compiler}--\S\ref{sec:eval}).
On successful admission, QBX emits an evidence-carrying IR containing
the normalized graph, execution schedule, type environment,
time-availability effects, linked kernels, and Merkle-style node and
root identities. The availability-effect system rejects
contract-visible look-ahead at compile time while admitting legal
point-in-time forecasts. Node identities support conditionally sound
incremental evaluation; the root identity, together with the evaluation
manifest, keys full-candidate deduplication and trial provenance.
Proposition~\ref{prop:coherence} states the resulting coherence
requirement precisely: internal cache hits are computational savings,
whereas only duplicate root programs are merged as trials.

(3)~\emph{A matched-budget evaluation of the benefits and costs of the
restricted space}
(\S\ref{sec:search}--\S\ref{sec:payoff}).
Under a common evaluator and matched search controls, evaluation-valid
proposals increase from 26.8\% for free strings and 36.4\% for a
grammar-only control to 97.7\% for typed programs; the additional
compiler checks account for 86.5\% of the measured admission gap
relative to that grammar control, and produce approximately $29\times$
more distinct valid factors over 300 iterations. Content identity enables
$7.50\times$ incremental replay with numerical parity and
positive held-out cross-lineage reuse in 15/15 cases. At matched LLM
budgets, typed amplification reaches the empirical best-reachable
frontier with at least $10\times$ fewer calls on the pre-extension
panel. The restriction also has a measurable boundary: typed-only books
lose behavioral breadth, while combining LLM-sourced breadth with
QBX-sourced depth improves test RankICIR by $+0.036$--$0.038$ in the
exploratory matched-budget study.
\section{Related Work}
\textbf{Alpha mining.} Formulaic-alpha generators span the canonical corpus \cite{kakushadze2016alphas}, evolutionary and learned graph search \cite{cui2021alphaevolve}, RL over expression construction \cite{yu2023alphagen,ren2024riskminer}, MCTS with syntax constraints \cite{shi2025alphajungle}, and LLM agents with retrieval, debate, or regularization \cite{wang2023alphagpt,tang2025alphaagent,guo2026alphaprobe,shi2026hubble,chen2026alphasage}. These systems innovate on the \emph{proposer}; admission, identity, and trial accounting remain bespoke per paper. Evaluation suites \cite{ding2026alphaeval} standardize scoring of finished factors but not the search-time layers where inflation enters. Shared infrastructure has venue precedent for \emph{other} pipeline layers---GPU limit-order-book simulators for RL trading \cite{frey2023jaxlob}---but no recent proceedings paper supplies the corresponding substrate for alpha mining itself.

\textbf{Backtest inflation and leakage.} That data-mined track records overstate out-of-sample performance is settled statistics: reality checks and SPA tests \cite{white2000reality,hansen2005spa}, the cross-section multiple-testing critique \cite{harvey2016crosssection}, probability of backtest overfitting \cite{bailey2016pbo}, deflated and haircut Sharpe ratios \cite{bailey2014deflated,harvey2015backtesting}, Sharpe inference under autocorrelation \cite{lo2002sharpe,ledoit2008robust}, and the evaluation discipline of asset-pricing ML \cite{gu2020ml}. Leakage has a parallel general-ML literature \cite{kaufman2012leakage,kapoor2023leakage}. All of it conditions on a \emph{known} trial set and a \emph{trusted} feature timeline; our substrate makes the trial count a system-recorded quantity and look-ahead a type error, upstream of those corrections.

\textbf{Programming-languages and systems lineage.} Every ingredient has precedent: strongly typed program search \cite{montana1995stgp}, graph-structured genetic programming \cite{miller2000cgp}, library learning \cite{ellis2021dreamcoder}, LLM proposers behind systematic evaluators \cite{romera2024funsearch}, equality saturation \cite{willsey2021egg}, term-rewriting foundations \cite{baader1998term}, build systems \cite{mokhov2018buildsystems}, and domain compilers \cite{lattner2021mlir}. Finance has platform-level strategy DSLs, and Qlib provides mature expression and dataset caching \cite{yang2020qlib}. What is new is the composition: one compiler contract that jointly checks legality, information timing, and identity for whole strategies, exposed as interfaces to search, evaluation, and selection.

\section{The Substrate, in Three Stages}\label{sec:substrate}

\subsection{The object: strategies as typed DAG, not texts}\label{sec:object}
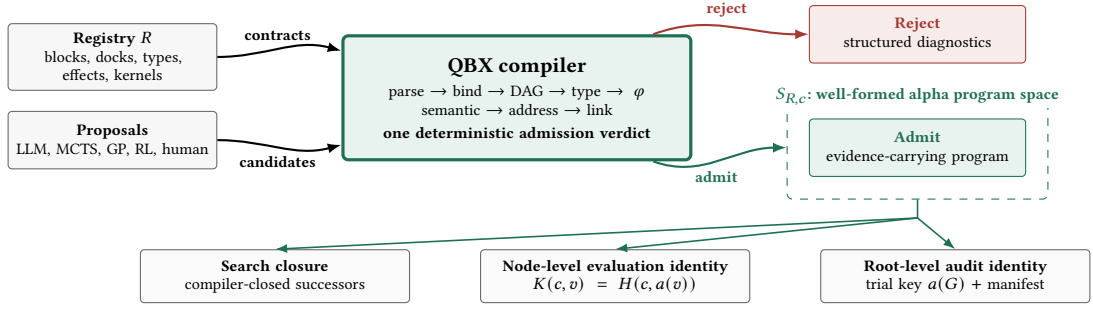
\begin{figure*}[t]
\centering
\begin{tikzpicture}[
  font=\scriptsize,
  input/.style={draw=black!60, rounded corners=1.5pt, fill=black!3,
    minimum height=0.76cm, text width=2.55cm, align=center, inner sep=3pt},
  compiler/.style={draw=qbxgreen, very thick, rounded corners=2pt, fill=qbxlight,
    minimum height=1.64cm, text width=4.25cm, align=center, inner sep=5pt},
  verdict/.style={draw=black!60, rounded corners=1.5pt,
    minimum height=0.76cm, text width=2.65cm, align=center, inner sep=3pt},
  contract/.style={draw=black!60, rounded corners=1.5pt, fill=black!2,
    minimum height=0.70cm, text width=3.35cm, align=center, inner sep=3pt},
  flow/.style={-{Latex[length=1.8mm]}, thick},
  contractflow/.style={-{Latex[length=1.6mm]}, semithick, draw=qbxgreen},
  lab/.style={font=\scriptsize\bfseries, fill=white, inner sep=1pt}
]
\node[input] (registry) at (0.90,0.72) {
  \textbf{Registry $R$}\\
  blocks, docks, types, effects, kernels
};
\node[input] (proposal) at (0.90,-0.38) {
  \textbf{Proposals}\\
  LLM, MCTS, GP, RL, human
};
\node[compiler] (compiler) at (6.25,0.18) {
  {\small\textbf{QBX compiler}}\\[2pt]
  parse $\to$ bind $\to$ DAG $\to$ type $\to\phieff$\\
  semantic $\to$ address $\to$ link\\[2pt]
  \textbf{one deterministic admission verdict}
};
\node[verdict, draw=rejectred, fill=rejectlight] (reject) at (11.55,1.02) {
  \textbf{\textcolor{rejectred}{Reject}}\\
  structured diagnostics
};
\node[draw=qbxgreen, dashed, rounded corners=2pt,
  minimum width=3.45cm, minimum height=1.36cm] (space) at (11.55,-0.48) {};
\node[verdict, draw=qbxgreen, fill=qbxlight] (admit) at (11.55,-0.48) {
  \textbf{\textcolor{qbxgreen}{Admit}}\\
  evidence-carrying program
};
\node[font=\scriptsize\bfseries, text=qbxgreen, fill=white, inner sep=1pt]
  at (11.55,0.20) {$S_{R,c}$: well-formed alpha program space};

\node[contract] (search) at (3.05,-2.18) {
  \textbf{Search closure}\\
  compiler-closed successors
};
\node[contract] (eval) at (7.55,-2.18) {
  \textbf{Node-level evaluation identity}\\
  $K(c,v)=H(c,a(v))$
};
\node[contract] (audit) at (12.05,-2.18) {
  \textbf{Root-level audit identity}\\
  trial key $a(G)$ $+$ manifest
};

\draw[flow] (registry.east) to[out=0,in=165]
  node[lab,pos=0.48,above=2pt]{contracts} (compiler.165);
\draw[flow] (proposal.east) to[out=0,in=195]
  node[lab,pos=0.48,below=2pt]{candidates} (compiler.195);
\draw[flow, draw=rejectred] (compiler.25) to[out=20,in=180]
  node[lab,pos=0.48,above=2pt,text=rejectred]{reject} (reject.west);
\draw[flow, draw=qbxgreen] (compiler.335) to[out=-20,in=180]
  node[lab,pos=0.48,below=2pt,text=qbxgreen]{admit} (space.west);
\coordinate (fan) at (11.55,-1.41);
\draw[draw=qbxgreen, semithick] (space.south) -- (fan);
\draw[contractflow] (fan) -- (search.north);
\draw[contractflow] (fan) -- (eval.north);
\draw[contractflow] (fan) -- (audit.north);
\end{tikzpicture}
\caption{QBX is the central contract boundary. Registry declarations and proposals enter the compiler; rejection emits structured diagnostics, while admission emits an evidence-carrying program in $S_{R,c}$. The same evidence supplies search closure, node-level evaluation identity, and root-level audit identity.}
\Description{Registry and proposal inputs feed a central QBX compiler. Rejected inputs produce diagnostics. Admitted evidence-carrying programs sit inside the well-formed alpha program space and feed three distinct consumers: search, node-level evaluation, and root-level audit.}
\label{fig:spine}
\end{figure*}

Let $T$ be the strings or trees miners emit and $S$ the strategies as \emph{functions} of market history. The denotation $D:T\rightharpoonup S$ is partial (most texts execute nothing) and many-to-one (one strategy has many spellings). Let $S_0\subset S$ contain strategies whose value at $t$ uses only information then available. Raw-text search exposes four obstructions: (O1)~invalid terms waste evaluation; (O2)~aliases are counted repeatedly; (O3)~scores cannot identify membership in $S_0$, because look-ahead can improve both train and test; and (O4)~evaluation is paid per term rather than reusable computation. A substrate is a computable presentation of $S$ that makes these properties checkable or priceable before evaluation; QBX uses a typed port-DAG. Figure~\ref{fig:spine} centers its compiler verdict and downstream contracts.

\subsection{Stage I: a registered strategy space}\label{sec:space}
A registry $R=(B,T_{\mathrm{val}})$ contains a finite set of versioned blocks (95 in the current registry) and a value-type vocabulary. Each block $b\in B$ declares named input and output \emph{docks} (typed ports), a parameter schema with defaults, a time-availability transfer function, a determinism flag, and an implementation entry point; its interface is a product signature $F_b:\prod_{i\in\mathrm{In}(b)}T_i\times\Theta_b\to\prod_{o\in\mathrm{Out}(b)}T_o$. A candidate strategy is a directed \emph{port graph} $G=(V,E,\lambda,\theta)$ whose edges connect a source node's output dock to a target node's input path; nodes carry roles (\textsf{source}, \textsf{compute}, \textsf{intent}), block references, and effective parameters. Sharing is explicit: one producer feeds many consumers without duplication.

\begin{table}[t]
\caption{Reference registry. Port counts sum declarations across block contracts; seven execution blocks make programs whole strategies.}
\label{tab:registry}
\small
\setlength{\tabcolsep}{4pt}
\renewcommand{\arraystretch}{0.94}
\begin{tabular}{@{}l r p{3.9cm}@{}}
\toprule
Surface & Count & Composition\\
\midrule
Node kinds & 3 & source, compute, intent\\
Block registry & 95 & 11 categories; 17 namespaces\\
Typed-port declarations & 261 & 155 inputs, 106 outputs; 2.75/block\\
Value-type vocabulary & 13 & expressions over 10 atomic types\\
Parameters & 83 & schema-checked slots, 32 names\\
Availability & 37 & annotated blocks; 39 temporal roles\\
\bottomrule
\end{tabular}
\end{table}

Table~\ref{tab:registry} makes the current admitted vocabulary and its representational boundary inspectable. The 261 figure counts named port declarations summed over 95 block contracts, not distinct port kinds: 155 inputs and 106 outputs, or 2.75 per block. Most contracts expose two or three ports; \texttt{ta.supertrend} is the largest, with three inputs and four outputs. The 95 contracts comprise 25 indicators, 13 series transforms, 12 operators, 12 risk rules, 10 edge detectors, 7 comparators, 7 execution models, 4 sizers, 3 logical gates, a temporal aggregator, and a timer; the seven execution components are why the admitted object is a whole strategy (signals, sizing, execution), not a factor expression. The declarations use only 13 type expressions over 10 atomic types; three use a wildcard type, confined to adapter boundaries. The dialect-reject accounting of \S\ref{sec:search} and the open representability census (\S\ref{sec:concl}) quantify that boundary.

\begin{defn}[Admitted space]\label{def:space}
For registry $R$ and compiler version $c$, let
$\WF_{R,c}(G) := \mathrm{Complete}(G)\wedge\mathrm{Bound}_R(G)\wedge\mathrm{DAG}(G)\wedge\mathrm{Typed}_R(G)\wedge\mathrm{Causal}_R(G)\wedge\mathrm{Semantic}(G)\wedge\mathrm{Linked}_R(G)$,
and $S_{R,c}=\{G \mid \WF_{R,c}(G)\}$: complete strategy documents that parse, bind only registered blocks, are acyclic, dock-type correct, time-availability safe, semantically well-shaped, and fully linked to executable kernels. Each conjunct is discharged by a named compiler phase.
\end{defn}

The space has the properties a search consumer needs. Its executable vocabulary is closed: nothing outside the registry can be referenced, so every admitted strategy is runnable by construction. Local dock-type checks cheaply prune candidate edits, while the whole-graph compiler remains authoritative because a port-compatible rewire can still create a cycle. Under a search profile of finite parameter domains and candidate caps, the one-step action set (Table~\ref{tab:actions}) is finite, and:

\begin{table}[t]
\caption{Compiler-closed graph actions. Every committed child is complete, boundary-compatible, and admitted; this is the ``homologous'' boundary in typed crossover (\ex{7}).}
\label{tab:actions}
\scriptsize
\renewcommand{\arraystretch}{0.94}
\begin{tabularx}{\columnwidth}{@{}p{1.55cm} >{\raggedright\arraybackslash}X@{}}
\toprule
Action & Legality condition and role\\
\midrule
Substitute & compatible named docks, parameter schema, and effect\\
Parameter & value from the profile's finite schema-valid domain\\
Reconnect & compatible non-descendant output; no cycle, no later information\\
Insert/delete & split an edge with a compatible block; bypass a node preserving every consumer\\
Clone/share & clone before specializing one consumer; merge only pure nodes with equal addresses\\
Subgraph edit & rewrite or cross over a region only under a matching typed/effect boundary\\
Typed hole & carry $(\tau,\phieff\text{-bound},\text{scope})$ until a compatible block or existing output fills it\\
\bottomrule
\end{tabularx}
\end{table}

\begin{prop}[Transition closure]\label{prop:closure}
$\WF_{R,c}(G)\wedge a\in A_{R,c,D}(G)\Rightarrow\WF_{R,c}(\mathrm{apply}(G,a))$. \emph{Sketch:} immediate by construction---the enumerator compiles each candidate child and returns it only on success; completeness of the action set over $S_{R,c}$ is bounded and not claimed.
\end{prop}

$S_{R,c}$ is therefore a \emph{definition space}: any proposer can target it, membership is a compiler verdict rather than an author's intent, and search never leaves the legal set. Raw text is a proposal, not a safe object.

\subsection{Stage II: a compiler for alpha mining}\label{sec:compiler}
Admission is a deterministic partial function $C_{R,c}:Y\rightharpoonup \mathrm{IR}$ over documents $Y$: ordinary CPU passes, no model inference, structured diagnostics on failure. Table~\ref{tab:obligations} lists what each phase rejects and what witness the IR persists: execution cannot rediscover a compiler-detectable fault, and downstream consumers read certificates rather than reinterpreting source text. This addresses O1: well-typedness, not a backtest crash, decides evaluability.

\begin{table}[t]
\caption{Compiler obligations and persisted IR evidence. The tuple $(\widehat G,\pi,\Gamma,\Phi,E_\tau,A_V,a(G),L)$ records the normalized graph through linked kernels.}
\label{tab:obligations}
\scriptsize
\renewcommand{\arraystretch}{0.94}
\begin{tabularx}{\columnwidth}{@{}p{1.28cm} >{\raggedright\arraybackslash}p{2.5cm} >{\raggedright\arraybackslash}X@{}}
\toprule
Phase & Rejects & Retained witness\\
\midrule
Normalize & malformed envelope; unknown block or dock; invalid parameter & normalized graph, effective parameters, registry provenance\\
Graph & unresolved reference; directed cycle & materialized edges, schedule $\pi$\\
Type & incompatible source and target docks & type environment $\Gamma$, typed edges\\
Availability & invalid temporal parameter; $\phieff>0$ & node and edge effects $\Phi$\\
Semantic & invalid intent boundary; unresolved kernel & intent closure, execution references $L$\\
Address & none beyond prior phases & node map $A_V$, root $a(G)$\\
\bottomrule
\end{tabularx}
\end{table}

\begin{defn}[Time-availability effect]\label{def:phi}
For a value with logical index $t$, $\phieff\in\mathbb{Z}\cup\{\bot\}$ is the signed offset at which the value becomes known: $\phieff=0$ at $t$, $\phieff<0$ earlier, $\phieff>0$ later (unavailable at $t$), $\bot$ for timeless literals. Multi-input joins take the maximum over timed inputs; registered transfers give $\phieff(\text{lag }p)=\phieff(x)-p$, $\phieff(\text{window }p)=\phieff(x)$ for $p>0$, $\phieff(\text{latency }l)=\mathrm{join}+l$. Admission requires every materialized node to have $\phieff\le 0$ or $\bot$.
\end{defn}

\begin{prop}[Relative non-anticipativity]\label{prop:phi}
Assume every source offset upper-bounds true publication time and every registered transfer soundly describes its kernel. If compilation succeeds, every node value at index $t$ depends only on information available by $t$. \emph{Sketch:} induction over the certified schedule; acceptance forces the join-plus-transfer bound $\le 0$ at every node and recursively checked inline expression.
\end{prop}

This addresses O3 \emph{without market data}. A value indexed $T{+}7$ but published at $T$ may carry $\phieff=-7$, so alignment by $-7$ is legal; the same shift on an ordinary $\phieff=0$ source yields $+7$ and is rejected. The guarantee remains relative to registered contracts: irregular timestamps, revisions, hindsight universe membership, corporate actions, and as-of joins remain in the trusted base. Zero admitted contract violations is therefore not a claim of zero backtest leakage.

\textbf{Measured leak evidence (\ex{2}).} Static checking rejected 6/6 planted look-ahead factors that a train/test split detected 0/6; an expanded battery rejected 10/10 adversarial temporal constructions and admitted 3/3 causal controls without loading a panel. Across 17 wild corpora, raw LLM output leaked at 0.17\% (2/1{,}151; 95\% binomial CI 0.02--0.63\%) and public-miner future reads at 0.10--0.20\%; checker and pipeline had zero disagreements across 3{,}628 co-scored rows. One public factor plausibly squared a one-period future return inside a volume--volatility term. Admission left 0/23{,}342 registered-contract violations. Incidence is generator-specific; the result shows that rare, contract-visible faults can be localized statically \cite{kaufman2012leakage,kapoor2023leakage}.

\textbf{Economic measurement and universe diagnostics (\ex{16}).} A warning-only extension attaches \emph{Unit}, \emph{Kind}, \emph{Scale}, \emph{Support}, and \emph{Aggregation} metadata plus universe/vintage/as-of provenance to source ports, then propagates them after dock typing and before availability. It localizes stock--flow arithmetic, unit-dependent $\log(\mathrm{market\ cap})$, and rank without point-in-time universe provenance without blocking admission, since basis-sensitive forms can be useful. In an isolated planted benchmark, the prototype localized 18/18 unambiguous faults, produced no hard error on 14 clean cases, and emitted 3/3 expected warnings: evidence of diagnostic representability, not alpha improvement.

\begin{defn}[Content identity]\label{def:addr}
Each node receives a Merkle-style address $a(v)$: the SHA-256 hash, over a canonical serialization with fixed float and NaN normalization, of its normalized executable semantics, effective parameters, declared output types, registered block contract, and its parents' addresses and selected docks. The graph root hashes the sorted multiset of node addresses together with compiler and registry versions. Authored names, positions, and presentation metadata are excluded.
\end{defn}

Three consequences are load-bearing: \emph{renaming invariance} (presentation edits preserve all addresses); \emph{cone-local invalidation} (an edit changes addresses only in its downstream dependency cone); and \emph{conservativeness}---equal addresses imply equal structural contracts, but $x$ and $x{+}0$ differ. Identity is a \emph{ladder} of coarser relations: exact string ${\subset}$ structural/canonical ${\subset}$ audited-rewrite saturation ${\subset}$ panel-extensional (equal evaluated values; never a cache key). The tiers are empirically distinct, and where each pays is itself a finding (\ex{3}, 30{,}210 arrivals): exact-string deduplication removes 40.1\% of raw LLM output but 0\% of post-admission search arrivals; one-shot canonicalization adds only 0.27\% there, bounded saturation 13.8\%; and 27.9\% of covered canonically distinct survivors are panel-extensional duplicates discovered only after evaluation---an honesty correction to breadth counts, never an evaluation saving. No single relation can play all four roles (O2).

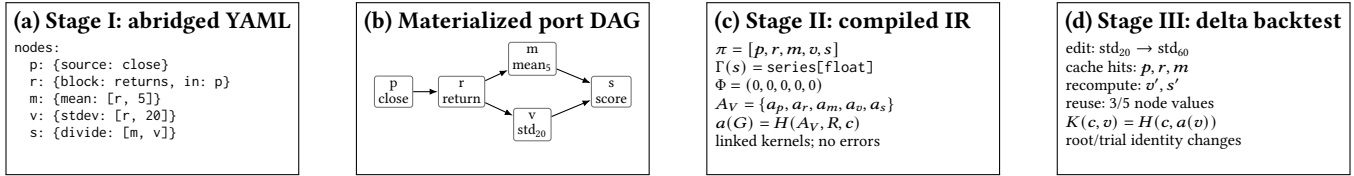
\begin{figure*}[t]
\centering
\setlength{\fboxsep}{3pt}
\fbox{\begin{minipage}[t][2.05cm][t]{0.205\textwidth}
\vspace{0pt}\textbf{(a) Stage I: abridged YAML}\par\vspace{2pt}
{\ttfamily\fontsize{5.8}{6.4}\selectfont
nodes:\par
\ \ p: \{source: close\}\par
\ \ r: \{block: returns, in: p\}\par
\ \ m: \{mean: [r, 5]\}\par
\ \ v: \{stdev: [r, 20]\}\par
\ \ s: \{divide: [m, v]\}\par}
\end{minipage}}\hfill
\fbox{\begin{minipage}[t][2.05cm][t]{0.205\textwidth}
\vspace{0pt}\textbf{(b) Materialized port DAG}\par\vspace{2pt}
\centering
\begin{tikzpicture}[x=0.72cm,y=0.58cm,
  n/.style={draw=black!55,rounded corners=1pt,inner sep=1.5pt,
    font=\fontsize{5.6}{6}\selectfont,align=center},
  e/.style={-{Latex[length=1.1mm]},thin}]
\node[n] (p) at (0,1.5) {p\\close};
\node[n] (r) at (1.25,1.5) {r\\return};
\node[n] (m) at (2.55,2.25) {m\\mean$_5$};
\node[n] (v) at (2.55,0.75) {v\\std$_{20}$};
\node[n] (s) at (4.0,1.5) {s\\score};
\draw[e] (p)--(r); \draw[e] (r)--(m); \draw[e] (r)--(v);
\draw[e] (m)--(s); \draw[e] (v)--(s);
\end{tikzpicture}
\end{minipage}}\hfill
\fbox{\begin{minipage}[t][2.05cm][t]{0.205\textwidth}
\vspace{0pt}\textbf{(c) Stage II: compiled IR}\par\vspace{2pt}
{\fontsize{6.2}{6.9}\selectfont
$\pi=[p,r,m,v,s]$\par
$\Gamma(s)=\texttt{series[float]}$\par
$\Phi=(0,0,0,0,0)$\par
$A_V=\{a_p,a_r,a_m,a_v,a_s\}$\par
$a(G)=H(A_V,R,c)$\par
linked kernels; no errors\par}
\end{minipage}}\hfill
\fbox{\begin{minipage}[t][2.05cm][t]{0.205\textwidth}
\vspace{0pt}\textbf{(d) Stage III: delta backtest}\par\vspace{2pt}
{\fontsize{6.2}{6.9}\selectfont
edit: $\mathrm{std}_{20}\to\mathrm{std}_{60}$\par
cache hits: $p,r,m$\par
recompute: $v',s'$\par
reuse: 3/5 node values\par
$K(c,v)=H(c,a(v))$\par
root/trial identity changes\par}
\end{minipage}}
\caption{End-to-end trace across the three stages. A YAML proposal materializes as a DAG with one return producer shared by two consumers; compilation retains schedule, types, availability, linkage, and addresses. Under an unchanged evaluation context, changing only the volatility window invalidates its downstream cone, so the backtest reuses $p,r,m$ and computes only $v',s'$. Addresses are abbreviated; cache soundness is conditional on Proposition~\ref{prop:cache}.}
\Description{Four panels show an abridged YAML strategy, its shared port DAG, the evidence-carrying compiler IR, and a delta backtest that reuses three unchanged node values after a volatility-window edit.}
\label{fig:e2e}
\end{figure*}

\textbf{Trace diagnostics.} The zero effect vector is not cosmetic: replacing the causal return input with a one-period future shift raises its affected path to $\phieff=+1$, so compilation rejects it before a panel loads. A robust-scale substitution is proposed only across a compatible typed/effect boundary and committed only when the resulting child compiles.

\subsection{Stage III: computation and backtesting}\label{sec:compute}
\begin{prop}[Conditional cache soundness]\label{prop:cache}
Key node values by $K(c,v)=H(\text{cache version},\ \text{context},\ a(v))$. If (i) the context key fixes data revision, universe, range, and execution semantics; (ii) the node and its ancestors are pure and deterministic (cacheability is ancestor-closed); (iii) registered contracts denote the deployed kernels; and (iv) collisions are negligible, then a cache hit equals fresh evaluation. \emph{Sketch:} topological induction---equal addresses identify equal contracts and equal parent addresses, hence equal inputs; purity gives equal outputs.
\end{prop}

This addresses O4: evaluation is paid per addressed computation, not per spelling. Compared with the occurrence ASTs commonly used for formula expressions, the QBX port DAG makes shared producers and stable subgraph identity first-class across a whole strategy and its local edits; it therefore captures reusable structure below surface syntax (Figure~\ref{fig:e2e}). Concretely, a $j$-stage feature feeding each stage to both time-series normalization and cross-sectional rank costs $T_j=4\cdot2^j-3$ nodes as an expanded tree but $D_j=3j+1$ as a port DAG. This is an execution-and-identity advantage, not extra expressive power: a let-bound, hash-consed AST can encode the same sharing (\S\ref{sec:eval}).

\textbf{Integrated identity contract.} A typed AST, memo table, and experiment tracker can reproduce individual components of the system. QBX's design claim is narrower: the Merkle scheme of Def.~\ref{def:addr} supplies one two-level identity algebra across them. \emph{Node} addresses key incremental computation inside a candidate (Prop.~\ref{prop:cache}), while the \emph{root} address, itself a function of node addresses, identifies the candidate and keys the trial record $r=(a(G),\ \text{registry manifest},\ c,\ \text{split},\ \text{metric},\ \text{selector},\ \text{proposer})$; trials with different data, selection, or evaluation semantics therefore remain distinct. A behavioral quotient (correlation-clustered niches; \S\ref{sec:payoff}) is a coarser relation derived from evaluated values and is used only for selection-side statistics, never for execution or deduplication.

\begin{prop}[Root-level ledger--cache coherence]\label{prop:coherence}
If full-candidate deduplication and trial accounting both key on the same root identity, an exactly duplicated candidate is one event in both ledgers: skipped by the evaluator and merged in the trial count. Node-level hits inside a novel candidate are cost savings, not trials. If deduplication and accounting key on different relations, the recorded trial count diverges from evaluation history by exactly the two partitions' disagreement set---spellings counted but never priced, or evaluations reused but never counted. \emph{Sketch:} both counters are pushforwards through the root partition; distinct partitions differ on exactly their disagreement cells.
\end{prop}

Because the root address is computed from node addresses, the computation and trial ledgers are projections of one algebra and cannot drift apart---a property not supplied by a memo table coupled to a separate tracker. The same coupling binds the other interfaces: admission must run on the normalized object the cache keys, or a leak verdict attaches to a different program than the one evaluated; closure needs the compiler \emph{inside} the action loop---a bolt-on validity filter is precisely the grammar arm of Table~\ref{tab:ladder}, and it pays the 61-point admission gap. Retro-fitting identity is measurably worse than owning it: a bolt-on AST cache stayed slower than no cache until ${\sim}1{,}100$ candidates while hash-consed-in identity reached $1.69\times$ (\ex{4}), and the identity ledger reproduces independently recounted denominators on 12/12 corpora only because both consume the same root-level canonical identities (\ex{3}).

\textbf{Rewrite audit (\ex{5}).} Algebraic identities require explicit semantic side conditions in a financial factor DSL. The exact tier terminates (0 cap hits in 12{,}000 random terms); the production structural tier has 0 idempotence or order-independence failures over 5{,}000 documents $\times$ 20 variants; and a 42{,}790-check collision audit is clean. By contrast, $x\cdot 0\to 0$ changes denotation on NaN/Inf panels (611/1{,}680), and unrestricted rule sets are order-dependent. Execution and accounting therefore key on exact structural identity, while audited saturation is restricted to an optional pre-evaluation quotient (removing 13.8\% of arrivals).

\textbf{Evidence protocol.} Table~\ref{tab:experiments} distinguishes panel, AAPL, and data-free studies. Panel rows use forward-adjusted BaoStock daily OHLCV, CSI500 primary and CSI300 where stated. The target is the 20-day forward return; rank IC is daily cross-sectional Spearman correlation and ICIR its time-series mean/standard deviation. Train ($\leq$2020-12-31) fixes signal direction, validation (2021-01--2022-06) selects, and test begins 2022-07; test-selected oracles are labeled diagnostics, not out-of-sample payoff. Pre-extension data end 2025-08; extended data end 2026-05 and use current-membership snapshots.

Rows \ex{1}/\ex{7}/\ex{8} and \ex{11} timings use 1{,}738 Yahoo Finance AAPL one-hour bars from 2025. The first two test validity, \ex{8} scores the first 60\%, and \ex{11} node counts are synthetic; none is cross-sectional evidence. Timings used a 10-core Apple M4 Mac mini with 24~GB RAM. A frozen \texttt{deepseek-chat} bank was generated 2026-06-11/12. Each call requested ten distinct CSI500 formulas over the stated OHLCV primitives and six fixed themes; its prompt targeted next-day return, while reported panel scores target 20 days. Temperature was 1.1, with a 4{,}096-token cap and three attempts. Confirmatory hypotheses and rules were internally preregistered.

\begin{table*}[t]
\caption{Experiment index. Each identifier denotes one study; detailed outcomes and limitations appear at the cited use. AAPL and data-free rows test mechanisms rather than panel-level predictive performance.}
\label{tab:experiments}
\centering
\fontsize{6.8}{7.4}\selectfont
\setlength{\tabcolsep}{3pt}
\renewcommand{\arraystretch}{0.96}
\begin{tabularx}{\textwidth}{@{}l >{\raggedright\arraybackslash}X l >{\raggedright\arraybackslash}X@{}}
\toprule
ID & Study and scope & ID & Study and scope\\
\midrule
\ex{1} & Typed vs. type-blind mutation; AAPL 1-hour executable yield. & \ex{9} & Matched-call frontier; CSI500, CSI300 check; test-selected diagnostic bound.\\
\ex{2} & Availability audit: planted faults/controls, 17 corpora, 23{,}342 admissions. & \ex{10} & Thirteen-action steelman string kernel; three exploratory seeds.\\
\ex{3} & Identity tiers over 30{,}210 arrivals and independent recounts on 12 corpora. & \ex{11} & AAPL 1-hour lineage replay; 1{,}760 parity checks; synthetic $N{=}246$ structures.\\
\ex{4} & Identity construction: bolt-on AST cache vs. address-at-construction. & \ex{12} & Held-out store prewarming across 15 search lineages.\\
\ex{5} & Randomized termination, confluence, collision, and NaN/Inf rewrite audit. & \ex{13} & LLM-only, typed-only, and hybrid books; $K{=}10/20/40$, three seeds.\\
\ex{6} & Six-arm representation ladder; extended CSI500, 10 seeds $\times$ 300 iterations. & \ex{14} & Occupancy-penalized selection over eight frozen, panel-scored pools.\\
\ex{7} & Homologous typed vs. unconstrained crossover; AAPL 1-hour validity. & \ex{15} & Common evaluator over frozen local-adapter pools; extended CSI500.\\
\ex{8} & Typed-hole vs. blind enumeration; AAPL 1-hour evaluations to optimum. & \ex{16} & Data-free economic-measurement and universe diagnostic battery.\\
\bottomrule
\end{tabularx}
\end{table*}

\section{Search: What Admission Buys at a Fixed Budget}\label{sec:search}
The primary search hypothesis is that typed admission improves candidate yield and measured quality at a fixed proposal budget. We compare against equal-prior string mutation, a grammar-only control, and a stronger 13-action string kernel so that syntactic filtering is not attributed to the type/effect system.

\begin{table}[t]
\caption{Fair representation ladder (\ex{6}; 6 arms, 10 seeds, 300 iterations, one evaluator; paired-seed 95\% CIs). ``Distinct'' is per run; ``Bank 20'' is iterations to 20 distinct factors. Typed arms A2/A3/A4 have bit-identical trajectories and differ only in cost instrumentation.}
\label{tab:ladder}
\scriptsize
\setlength{\tabcolsep}{2.4pt}
\renewcommand{\arraystretch}{0.95}
\begin{tabular}{lrrrl}
\toprule
Arm & Valid & Distinct & Bank 20 & Quality ($\Delta$ val.\ ICIR)\\
\midrule
A0 string$+$priors & 26.8\% & 7.1 & never & degenerate (bare leaf 10/10)\\
A1 grammar, same priors & 36.4\% & 38.0 & $140{\pm}28$ & $+0.140$ vs.\ A0 $[0.03,0.25]$\\
A2/A3 typed tree/DAG & 97.7\% & 204.9 & $20.5{\pm}0.5$ & $+0.132$ vs.\ A1 $[0.02,0.25]$\\
A4 $=$A3$+$memo & 97.7\% & 204.9 & $20.5{\pm}0.5$ & $=$A3; $-36.6\%$ node evals\\
A5 typed, priors off & 98.1\% & 202.0 & $20.5{\pm}0.5$ & $+0.041$ vs.\ A3 $[0.01,0.07]$\\
\bottomrule
\end{tabular}
\end{table}

\textbf{Fair representation ladder (\ex{6}, extended panel).} High typed-arm validity follows from compiler closure (Prop.~\ref{prop:closure}); the controls identify which obligations account for the difference. Grammar legality explains 13.5\% of the admission gap ($+9.6$pp, CI $[8.4,10.8]$), while type/effect admission explains 86.5\% ($+61.3$pp, CI $[60.0,62.6]$). The residual 2.3\% compiles but fails data-dependently during evaluation because of degenerate series or NaN panels. Admission does not ensure factor diversity: distinct valid factors nevertheless rise $7\to38\to205$, and iterations required to bank 20 fall from never to $20.5\pm0.5$. Disabling priors improves the typed arm by $0.041$, so priors do not explain the result. The decomposition remains relative to the production-rule grammar control; a type-constrained decoder could shift it. External-corpus dialect rejects are 13--26\% on the fullest typed pools, and a public-formula representability census remains open (\S\ref{sec:concl}). At this scale transpositions merge no candidates in 18{,}000 iterations, while memoization avoids 36.6\% of node evaluations but adds 16.9\,s of wall time; avoided evaluations, rather than wall time, are the relevant measure for these cheap kernels.

\textbf{Corroboration} (AAPL one-hour mechanism setting): homologous typed mutation yields 93.1\% executable candidates versus 9.7\% type-blind (\ex{1}); typed crossover 94.9\% versus 33.2\% valid offspring (\ex{7}); typed-hole guidance reaches the same optimum in 612 evaluations versus 5{,}220 for blind enumeration (\ex{8}).

\begin{figure*}[t]
\centering
\includegraphics[width=0.98\textwidth]{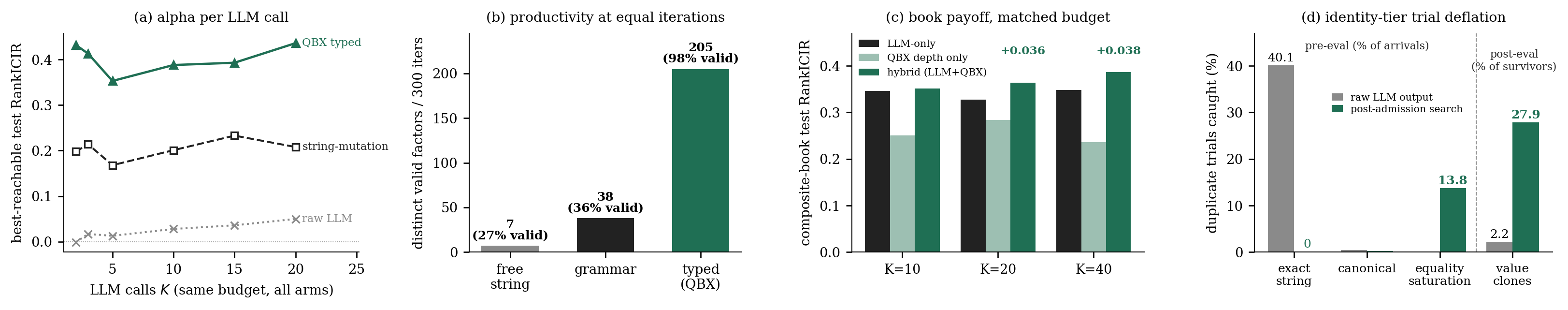}
\caption{QBX at matched budget. (a)~Best-reachable test RankICIR vs.\ LLM calls on extended CSI500 (\ex{9}; oracle is a test-window-selected diagnostic upper bound). (b)~Distinct valid factors and validity rates (\ex{6}). (c)~Across three exploratory seeds, typed depth trails LLM breadth, while the hybrid recovers both (\ex{13}). (d)~Exact-string identity catches 40.1\% of raw LLM output but 0\% after admission; saturation removes 13.8\% pre-evaluation and the ledger finds 27.9\% value clones post-evaluation (\ex{3}).}
\Description{Four panels show RankICIR versus LLM calls, valid-factor productivity, matched-budget book payoff, and duplicate-trial rates across identity tiers.}
\label{fig:main}
\end{figure*}

\textbf{Compute-to-alpha frontier (\ex{9}).} Because admission and deduplication cost 56.7\,$\mu$s per candidate against a measured 1.45\,s median LLM call, each call is amplified into ${\sim}400$ legal candidates of CPU-bound search. Figure~\ref{fig:main}(a) shows the consequence on the extended panel: the typed arm's best-reachable RankICIR sits at 0.35--0.44 from $K{=}2$ calls onward, roughly double the string-mutation control at every budget, while raw LLM sampling stays near zero; on the pre-extension panel the typed arm reaches the panel's reachable ceiling from $K{=}2$--3 calls, a ${\ge}10\times$ call reduction at matched quality (oracle $=$ test-window-selected upper bound, a diagnostic). The oracle edge over string mutation is $+0.10$/$+0.12$ at $K{=}10/20$ and vanishes at $K{=}40$; on CSI300 it weakens (mean $+0.061$). An internally preregistered \emph{steelman} string kernel (13 mutation actions; \ex{10}; exploratory at $n{=}3$ seeds) closes the typed oracle gap at $K{=}10$ and leaves the typed arm ahead but not seed-significant at $K{=}20$: the robust distinctions are admission, cost, and contract-checked timing, and we scope the generation claim accordingly.

\textbf{One evaluator, many miner adapters (\ex{15}).} Local reimplementations of public miners are compiled under one evaluator, deduplication funnel, and selection protocol (Table~\ref{tab:funnel}). The funnel exposes multiplicity and partiality: the Hubble adapter's 1{,}000 emissions reduce to 250 canonical identities, while 33\% of the AlphaAgent adapter's unconstrained expressions remain unscoreable. The QBX-native arm compiles 1{,}000 complete strategies at 100\% admission, and QBX--Hubble routes Hubble-style proposal logic through the registered surface. Because these are local adapters without official-code fidelity checks, the study demonstrates interoperability and common accounting rather than ranking the published methods.

\begin{table*}[t]
\caption{One evaluator, many miners (\ex{15}; extended CSI500). Exact/Canon.\ are distinct texts/identities; Scored have complete metrics; Sem.\ is static validity; $t_k$ is test RankICIR of validation-selected top $k$. Adapters are local and not official-code fidelity checks; one frozen pool is used per arm. This is an interoperability demonstration, not a method ranking.}
\label{tab:funnel}
\centering
\scriptsize
\setlength{\tabcolsep}{4.5pt}
\renewcommand{\arraystretch}{0.95}
\begin{tabular}{@{}l rrrrr rrrr@{}}
\toprule
Arm & Pool & Exact & Canon. & Scored & Sem. & $t_1$ & $t_5$ & $t_{10}$ & $t_{20}$\\
\midrule
QBX-factor adapter & 1{,}000 & 516 & 454 & 430 & 95.8\% & .460 & .442 & .428 & .417\\
Hubble adapter & 1{,}000 & 256 & 250 & 249 & 99.9\% & .454 & .419 & .390 & .345\\
QBX--Hubble adapter & 717 & 179 & 179 & 178 & 99.4\% & .277 & .338 & .315 & .288\\
AlphaAgent adapter & 1{,}000 & 759 & 754 & 507 & 67.0\% & .287 & .311 & .302 & .283\\
QBX-native$^\dagger$ & 1{,}000 & 1{,}000 & 1{,}000 & 1{,}000 & 100\% & .232 & .291 & .299 & .302\\
AlphaPROBE adapter & 1{,}000 & 312 & 312 & 279 & 94.9\% & .332 & .327 & .281 & .272\\
\bottomrule
\end{tabular}

\raggedright\scriptsize $^\dagger$Complete strategy graphs (signals, sizing, execution), not continuous factors: object-coverage evidence, not a factor-ranking row.
\end{table*}

\section{Evaluation: Identity-Priced Reuse}\label{sec:eval}
Search lineages are mostly re-computation: a child strategy differs from its parent in one dependency cone. Content identity turns that observation into a contract---equal addresses may be reused, everything else must be recomputed---and Prop.~\ref{prop:cache} prices the reuse. Table~\ref{tab:econ} summarizes.

\begin{table}[t]
\caption{Identity-priced evaluation economics. Wall-clock results are machine-dependent; node-evaluation ratios are structural.}
\label{tab:econ}
\footnotesize
\setlength{\tabcolsep}{2pt}
\renewcommand{\arraystretch}{1.04}
\begin{tabularx}{\columnwidth}{@{}>{\raggedright\arraybackslash}X
  >{\raggedright\arraybackslash}p{0.17\columnwidth}
  >{\raggedleft\arraybackslash}p{0.39\columnwidth}@{}}
\toprule
Measurement & Case & Result\\
\midrule
Incremental evaluation (\ex{11}) & wall & $7.50\times$\\
 & parity & 0/1{,}760 mismatches\\
\addlinespace[2pt]
Node-evaluation reduction ($N{=}246$) & tree & $21.4\times$\\
 & wide DAG & $48.4\times$\\
\addlinespace[2pt]
Inner loop (mutate$+$check$+$dedup) & candidate & 56.7\,$\mu$s\\
 & vs.\ LLM & $2.6{\times}10^{3}$--$1.3{\times}10^{5}\times$\\
\addlinespace[2pt]
Store replay (15 arms) & cold & $7.35\times$\\
 & prewarmed & $8.08\times$\\
\addlinespace[2pt]
Held-out prewarm (\ex{12}) & positive & 15/15 arms\\
 & median lift & $+0.011$\\
\bottomrule
\end{tabularx}
\end{table}

\textbf{Realized economics (\ex{11}).} Replaying a real search lineage on the AAPL one-hour series, incremental evaluation was $7.50\times$ faster than full re-evaluation with zero value mismatches in 1{,}760 incremental-versus-full comparisons; node-evaluation reductions reached $21.4\times$ (tree families) and $48.4\times$ (wide DAG families) on synthetic graphs of $N{=}246$ nodes (the ratios are structural, the wall clock is machine-dependent). The inner loop costs 56.7\,$\mu$s per candidate against a 1.45\,s median LLM call---between $2.6{\times}10^{3}\times$ and $1.3{\times}10^{5}\times$ depending on amortization (one call yielding ten candidates at the low end; five calls per kept candidate at the high end)---which is why admission belongs before, not inside, the proposer. Pooled store replay over 15 held-out lineage arms ran with $7.35\times$ (cold store) and $8.08\times$ (prewarmed) fewer node evaluations, per-arm prewarmed ratios $4.2$--$12.0\times$.

\textbf{Cross-lineage reuse, with its boundary.} Within-lineage reuse compounds in 8/8 arms; across lineages it is real but distance-dependent---no single global store dominates every specialist store, and prewarming lifts late-half reuse in 15/15 held-out lineages (median $+0.011$, max $+0.063$) with magnitude tracking kernel-family proximity (\ex{12}). Reuse is not a free lunch: as \S\ref{sec:search}'s memoization accounting shows, bookkeeping can dominate cheap kernels at small scale.

\textbf{Scope versus Qlib.} Qlib ships mature expression, memory, and dataset caches keyed by hashed query content \cite{yang2020qlib}; content hashing is not the novelty. The substrate's distinction is the \emph{joint scope} of the identity: compiler-checked, typed, time-availability-audited whole-strategy nodes, with one dependency identity serving execution reuse, graph transactions, and duplicate control. That is an object-model claim; no matched Qlib benchmark exists in our experiments, and we make no performance-ranking claim against it. A hash-consed AST with let-bound sharing is the correct strong baseline and remains future work; until it runs, we cannot attribute the $7.5\times$ among content identity, plain memoization, and evaluator implementation. One further stated gap: the cache key binds registered contracts, not implementation digests---kernel code changed without a version bump defeats it.

\section{Predictive Utility Under a Restricted Search Space}\label{sec:payoff}
$S_{R,c}$ is a strict subset of the free-form text space sampled by an LLM: the 95-block registry, typed docks, and time-availability constraints exclude some expressible programs by design. We therefore evaluate the restriction empirically rather than assume that it preserves predictive opportunity. Figure~\ref{fig:main} separates two effects: candidate-level reachability within the observed pools and book-level breadth across selected candidates.

\textbf{Candidate-level reachability.} At equal LLM-call budgets, the typed arm's best observed RankICIR is approximately twice that of the basic string-mutation control and matches the pre-extension pool's diagnostic frontier from $K{=}2$--3 calls (Figure~\ref{fig:main}a; \S\ref{sec:search}). Because that frontier is selected on the test window and registry coverage is incomplete, this result supports lower search cost within the evaluated pools, not a general claim that the restriction is costless. Compiler admission costs $2.6{\times}10^{3}$--$1.3{\times}10^{5}$ times less per candidate than the measured LLM-call baseline.

\textbf{Book-level payoff (\ex{13}).} Typed-depth search alone trails LLM-only books at every budget (test RankICIR 0.251/0.283/0.236 vs.\ 0.316/0.321/0.343 at $K{=}10/20/40$; three exploratory seeds). Effective breadth is 1.00--1.93 for typed search versus 4.65--9.38 for LLM proposals, and two typed composites are numerically identical to four decimals after sign alignment. Combining LLM-sourced breadth with typed search depth yields $+0.036\pm0.011$ over LLM-only at $K{=}20$ and $+0.038\pm0.010$ at $K{=}40$ (3/3 seeds; $+0.006$ at $K{=}10$, within seed noise; Figure~\ref{fig:main}c). The evidence therefore indicates complementarity between proposal breadth and compiler-closed local search in these pools, while also documenting a measurable breadth cost for typed-only search.

\textbf{Compiler-induced regularization (\ex{3}, \ex{6}).} Admission restricts the hypothesis class before evaluation, and \ex{6}/\ex{9} compare that restriction with non-typed generation. Identity controls multiplicity: the same quotient used for reuse also adjusts the trial count (Figure~\ref{fig:main}d; alias inflation is 1.2--1.64$\times$ on mutation-heavy lineages). Exact-string identity removes 0\% of post-admission search arrivals and one-shot canonicalization 0.27\%; recovering the remaining structural duplicates requires compiler-equivalent normalization. Post-hoc behavioral controls remain portable: niche-occupancy penalties reduce selected-basket mean $|\mathrm{corr}|$ from 0.740 to 0.253 and raise $N_{\mathrm{eff}}$ from 1.66 to 5.49 (\ex{14}), but recover at most 36\% of the oracle--naive selection gap. The regularization claim therefore rests on compile-time capacity and multiplicity control rather than the post-hoc selector.

\section{Conclusion and Future Work}\label{sec:concl}
\textbf{Discussion.} The trusted base is explicit: provider timestamps, kernel purity, and context-key completeness are assumed, not verified. Evidence is single-market (two Chinese A-share universes, 20-day horizon, pre-cost cross-sectional statistics); the extended vintage uses a current-membership universe snapshot. The steelman and book studies use three seeds (exploratory); leak-incidence rates are generator-specific; the mechanism probes are AAPL-only; and the LLM prompt's next-day wording differs from the 20-day scoring horizon. No matched Qlib or hash-consed-AST benchmark exists, and the latter is the correct strong baseline for the identity claims; several proposed graph actions remain prototypes; no representability census against public formula corpora exists, and typed-book mode collapse may partly reflect action-set coverage rather than proposer behavior alone. Finance realism---costs, capacity, execution---is out of scope.

\textbf{Future work.} The trial ledger makes selection auditing well-posed: root-deduplicated counts and behavioral-niche denominators plug into deflation statistics. Preliminary calibration finds the field-standard deflated-Sharpe construction anti-conservative at overlapping horizons; joint max-statistic bootstraps \cite{white2000reality,hansen2005spa} and adaptive-search-aware inference deserve separate study. Next steps are hash-consed-AST and Qlib benchmarks, a public-formula representability census, type-constrained decoding, and second-market replication.

\textbf{Conclusion.} We presented \emph{QBX}, a compiler that defines a well-formed, auditable alpha program space. Compiler admission checks structure, types, information availability, semantics, and executable linkage, then emits an evidence-carrying IR whose node identities support incremental evaluation and whose root identity governs candidate deduplication and trial accounting (Prop.~\ref{prop:coherence}). Across Chinese A-share panels, evaluation validity rises from 26.8\% to 97.7\%; the availability effect rejects tested contract-visible look-ahead; and incremental replay is $7.5\times$ faster with $0/1{,}760$ value mismatches. At matched LLM-call budgets, typed search reaches the pool-specific diagnostic frontier with ${\ge}10\times$ fewer calls on the pre-extension panel, while a breadth/depth hybrid improves test RankICIR by $+0.036$--$0.038$. QBX does not manufacture alpha; its contribution is to make alpha programs well-formed, reusable, and auditable for compatible miners.

\bibliographystyle{ACM-Reference-Format}
\bibliography{references}

\end{document}